\documentclass[prb,twocolumn,showpacs,superscriptaddress]{revtex4}
\usepackage{ulem}
\usepackage{amssymb}
\usepackage{subfigure}
\usepackage{graphicx}
\usepackage{amsmath}
\usepackage{amsfonts}

\begin{document}

\title{Theory for superconductivity in alkali chromium arsenides A$_2$Cr$_3$%
As$_3$ (A=K,Rb,Cs)}
\date{\today }
\author{Yi Zhou}
\affiliation{Department of Physics, Zhejiang University, Hangzhou 310027, China}
\affiliation{Collaborative Innovation Center of Advanced Microstructures, Nanjing 210093,
China}
\author{Chao Cao}
\affiliation{Department of Physics, Hangzhou Normal University, Hangzhou 310036, China }
\author{Fu-Chun Zhang}
\affiliation{Department of Physics, Zhejiang University, Hangzhou 310027, China}
\affiliation{Collaborative Innovation Center of Advanced Microstructures, Nanjing 210093,
China}

\begin{abstract}
We propose an extended Hubbard model with three molecular orbitals on a hexagonal lattice with $D_{3h}$ symmetry to study recently discovered
superconductivity in A$_2$Cr$_3$As$_3$ (A=K,Rb,Cs). Effective pairing interactions
from paramagnon fluctuations are derived within the random phase approximation, and are found to be most attractive in spin triplet channels.
At small Hubbard $U$ and moderate Hund's coupling, the pairing arises from 3-dimensional (3D) $\gamma$ band and has a spatial symmetry $f_{y(3x^{2}-y^{2})}$, 
which gives line nodes in the gap function. At large $U$, a fully gapped $p$-wave state, $p_{z}\hat{z}$ dominates at the quasi-1D $\alpha $-band.
\end{abstract}

\pacs{74.20.-z; 74.20.Mn; 74.20.Rp; 74.70.-b}
\maketitle

\begin{figure}[tbp]
\subfigure[Top View]{
   \includegraphics[width=4cm]{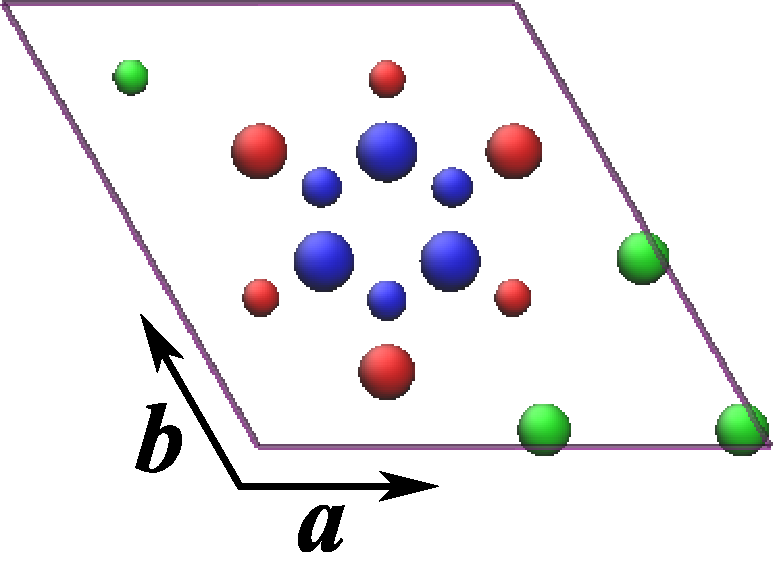}\label{fig1a}} 
\subfigure[(CrAs)$_6$ SNT]{
   \includegraphics[width=4cm]{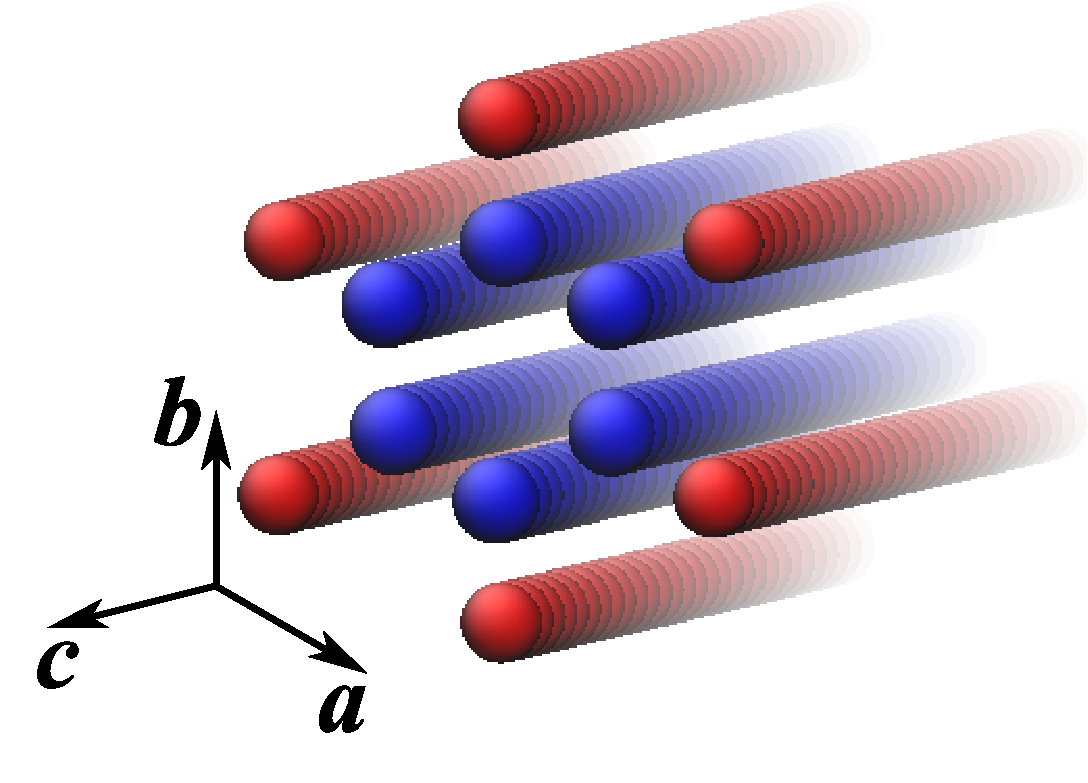}\label{fig1b}}
\caption{Crystal structure of $A_{2}$Cr$_{3}$As$_{3}$. (a) Top view and (b)
the (Cr$_6$As$_6$)$_{\infty}$ sub-nanotube. In panel (a), larger and smaller
atoms are at $z$=0.5 and $z$=0, respectively. In both panels, green balls
indicate alkaline atoms, red balls arsenic atoms, and blue balls chromium
atoms.}
\label{fig1}
\end{figure}

Recently, CrAs based superconductors have attracted much attention. CrAs
itself is a 3D antiferromagnet (AFM), which becomes superconducting (SC)
under modest pressure with $T_{c}\sim 2$K\cite{Luo14}. Subsequently, a new
family of quasi-1D superconductors A$_{2}$Cr$_{3}$As$_{3}$ (A=K,Rb,Cs) has
been discovered at ambient pressure with $T_c$ up to $6.1$K\cite%
{BaoK,TangRb,TangCs}. The key building block of A$_{2}$Cr$_{3}$As$_{3}$ is
the 1D [(Cr$_{3}$As$_{3}$)$^{2-}$]$_{\infty }$ double-walled sub-nanotubes
(Fig. \ref{fig1}), which are separated by columns of K$^{+}$ ions, in
contrast to the layered iron-pnictide and copper-oxide high Tc
superconductors.

This new family exhibits interesting properties in both the normal and SC
states. In the normal state, the resistivity follows $\rho (T)=\rho _{0}+AT$
in a wide temperature region, different from the usual Fermi liquid behavior 
$\rho _{0}+AT^{2}$\cite{BaoK,TangRb,TangCs}. NMR measurements on K$_{2}$Cr$%
_{3}$As$_{3}$ showed a power-law temperature dependence $1/T_{1}\sim
T^{0.75} $ above $T_{c}$, which is neither $1/T_{1}\sim T$ for a Fermi
liquid nor Curie-Weiss behavior $1/T_{1}T\sim C/(T+\theta )$ for an AFM\cite%
{Imai15}. Below $T_{c}$, the electronic contribution to the specific heat $%
C_{e}(T)$ deviates from the BCS senario, and the extrapolated upper critical
field $H_{c2}$ exceeds the Pauli limit\cite{BaoK,TangRb,TangCs}. The
Hebel-Slichter coherence peak of $1/T_{1}$ is absent in K$_{2}$Cr$_{3}$As$%
_{3}$\ NMR measurement\cite{Imai15}. London penetration depth measurement
for K$_{2}$Cr$_{3}$As$_{3}$ shows linear temperature dependence, $\Delta
\lambda (T)\sim T$, at temperatures $T\ll T_{c}$\cite{Yuan15}. All these
experiments are very difficult to explain within electron-phonon coupling
mechanism, and suggest unconventional nature of superconductivity.

The electronic structure of K$_{2}$Cr$_{3}$As$_{3}$ has been investigated by
Jiang \textit{et al.}\cite{Cao14} using density functional theory (DFT),
which is confirmed by later calculation\cite{HuJP15}. There are three energy
bands at the Fermi level: two quasi-1D $\alpha $- and $\beta $-bands with
flat Fermi surfaces, and a 3D $\gamma $-band. It is natural to ask the
question, which band is responsible for the superconductivity? On one hand,
the linear temperature dependent resistivity and the power-law $1/T_{1}$ in
NMR indicate a quasi-1D Tomonaga-Luttinger liquid; on the other hand, upper
critical field measurement $H_{c2}$ implies a 3D superconductor\cite%
{Canfield15}. In this Letter, we shall carry out theoretical study to
address this issue. We shall focus on K$_{2}$Cr$_{3}$As$_{3}$ and the theory
may be extended to other alkali chromium arsenides.

We begin with constructing the Hamiltonian by using symmetries. The space
group for A$_{2}$Cr$_{3}$As$_{3}$ lattices is $P\bar{6}m2$ and the
corresponding point group is $D_{3h}$.\cite{BaoK,TangRb,TangCs} We shall
also assume that the time reversal symmetry remains unbroken, and consider a
system described by the Hamiltonian%
\begin{equation*}
H=H_{0}+H_{int},
\end{equation*}%
where $H_{0}$ is the non-interacting part and $H_{int}$ is the interacting
Hamiltonian.

\emph{Tight-binding model.} We assume $H_{0}$ to be given by a tight-binding
Hamiltonian. For K$_{2}$Cr$_{3}$As$_{3}$, there are three Fermi surfaces
corresponding to the $\alpha $, $\beta $ and $\gamma $ bands. Minimally, we
need three orbitals per unit cell (per Cr$_{6}$As$_{6}$ cluster). From the
DFT calculation, there are three low energy molecular orbitals. Two of them
belong to 2D irreducible representation $E^{\prime }$ of $D_{3h}$ group, and
the other one is in 1D representation $A_{1}^{\prime }$. We denote two $%
E^{\prime }$ states as $\left\vert 1\right\rangle $ and $\left\vert
2\right\rangle $ and the $A_{1}^{\prime }$ state as $\left\vert
3\right\rangle $. Neglecting spin-orbit coupling, the tight-binding
Hamiltonian $H_{0}$ can be constructed from these three orbitals,%
\begin{equation}
H_{0}=\sum\limits_{\mathbf{k}mns}c_{\mathbf{k}ms}^{\dagger }\xi _{\mathbf{k}%
mn}c_{\mathbf{k}ns},  \label{H0}
\end{equation}%
where $m,n=1,2,3$ are the three molecular orbitals, $s=\uparrow ,\downarrow $
is the spin index, $c_{\mathbf{k}ms}^{\dagger }(c_{\mathbf{k}ms})$ creates
an $m$-orbital electron with spin $s$. In the basis $\left\{ \left\vert
1\right\rangle ,\left\vert 2\right\rangle ,\left\vert 3\right\rangle
\right\} $, $\xi _{\mathbf{k}mn}$ can be written in terms of matrix form%
\begin{equation}
\hat{\xi}_{\mathbf{k}}=\sum_{\tau =0}^{8}\xi _{\mathbf{k}}^{\tau }\lambda
_{\tau },  \label{H01}
\end{equation}%
where $\lambda_{0}$ is the unit matrix and $\lambda _{1-8}$ are Gell-Mann
matrices.

\begin{figure}[h]
\caption{Comparison of the Fermi surfaces for $K_2Cr_3As_3$ obtained from
the DFT calculation (a), and from the present model Hamiltonian (b). The $%
\protect\alpha$, $\protect\beta$ and $\protect\gamma$ bands are labelled.
The $\Gamma$ point is at the center of the Brillouin zone. }
\label{fig.2}%
\subfigure[]{
   \includegraphics[width=2.4cm]{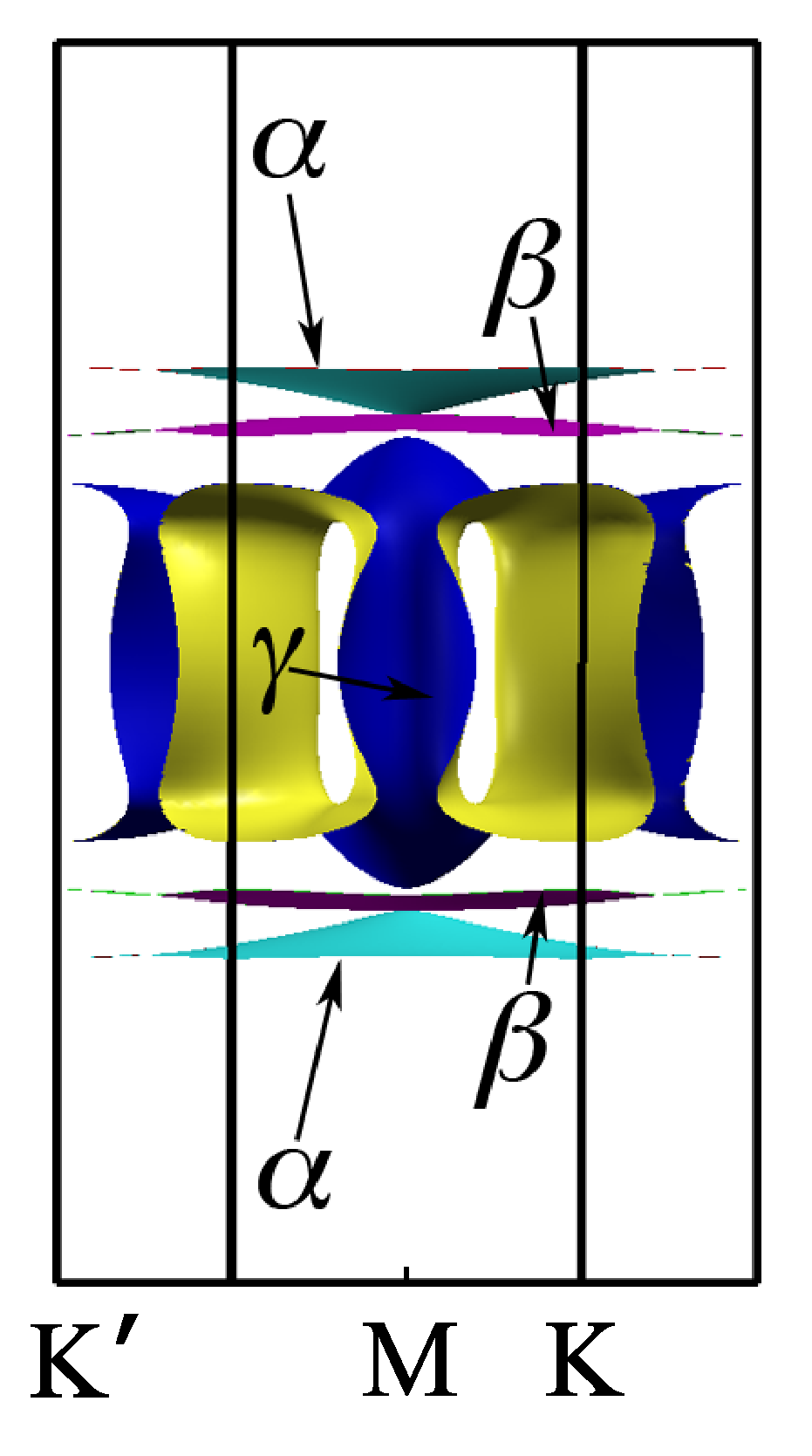}\label{fig2a}} \hspace{1cm} 
\subfigure[]{
   \includegraphics[width=2.4cm]{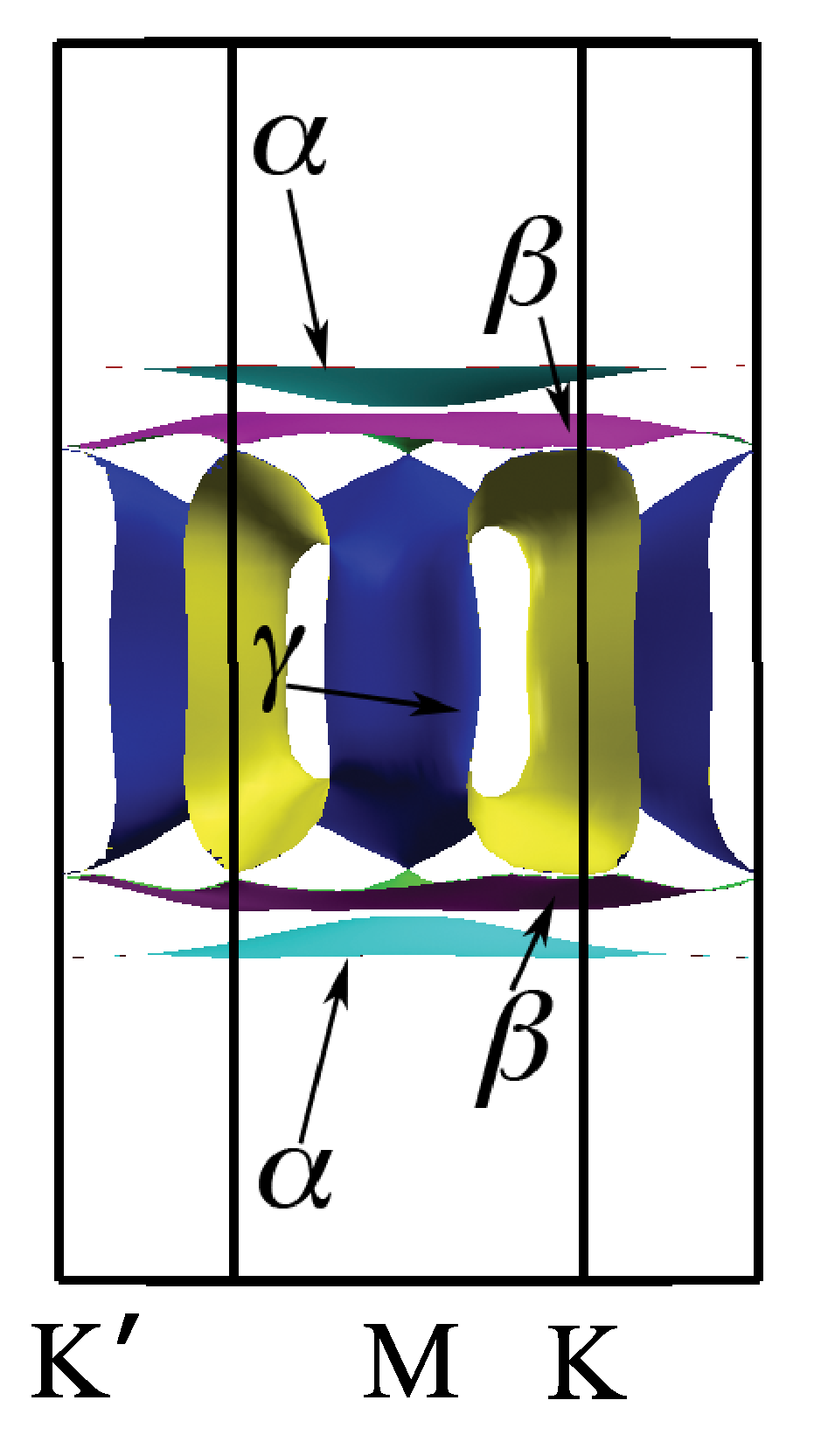}\label{fig2b}}
\end{figure}

We now use the symmetry to analyse $\xi _{\mathbf{k}}^{\tau },\tau =1,\cdots
,8$. $\lambda _{1-8}$ transfer as irreducible representations under $D_{3h}$
symmetry operations. Namely, $\lambda _{0}$ and $\lambda _{8}$ transfer as $%
A_{1}^{\prime }$, $\lambda _{2}$ transfers as $A_{2}^{\prime }$, $(\lambda
_{1},\lambda _{3})$, $\left( \lambda _{4},\lambda _{6}\right) $ and $\left(
\lambda _{5},\lambda _{7}\right) $ transfer as $E^{\prime }$ respectively. $%
H_{0}$ should be invariant under all the $D_{3h}$ symmetry operations,
thereby belongs to representation $A_{1}^{\prime }$. Therefore, $\xi _{%
\mathbf{k}}^{\tau }$ can be determined using the Clebsch-Gordan coefficients.%
\cite{Butler} Since the $ab$-plane lattice constant $a=9.98$\AA\ is much
larger than that along the $c$-axis $c=4.23$\AA , we will only keep the
hopping terms on $ab$-plane up to the first nearest neighbor (NN) bonds and
those along $c$-axis up to the second NN bonds. To do this, we set the
Bravais lattice basis $\mathbf{a}=(\frac{a}{2},-\frac{\sqrt{3}a}{2},0),%
\mathbf{b}=(\frac{a}{2},\frac{\sqrt{3}a}{2},0),\mathbf{c}=(0,0,c)$ (see
Fig.1), and introduce $k_{a}=\mathbf{k}\cdot \mathbf{a},k_{b}=\mathbf{k}%
\cdot \mathbf{b},k_{c}=\mathbf{k}\cdot \mathbf{c}$, and some harmonic
functions on the $ab$ plane,%
\begin{eqnarray}
s_{x^{2}+y^{2}}(\mathbf{k}) &=&\cos k_{a}+\cos k_{b}+\cos (k_{a}+k_{b}), 
\notag \\
p_{x}(\mathbf{k}) &=&2\sin (k_{a}+k_{b})+\sin k_{a}+\sin k_{b},  \notag \\
p_{y}(\mathbf{k}) &=&\sqrt{3}(\sin k_{b}-\sin k_{a}),  \label{HF1} \\
d_{x^{2}-y^{2}}(\mathbf{k}) &=&\cos k_{a}+\cos k_{b}-2\cos (k_{a}+k_{b}), 
\notag \\
d_{xy}(\mathbf{k}) &=&-\sqrt{3}(\cos k_{b}-\cos k_{a}).  \notag
\end{eqnarray}%
It is easy to verify that $s_{x^{2}+y^{2}}\propto 1-\frac{1}{4}%
(k_{x}^{2}+k_{y}^{2})$, $p_{x}\propto k_{x}$, $p_{y}\propto k_{y}$, $%
d_{x^{2}-y^{2}}\propto k_{x}^{2}-k_{y}^{2}$ and $d_{xy}\propto k_{x}k_{y}$
at small $ka$. Thus, we have $\xi _{\mathbf{k}}^{2}=0$, and other nonzero $%
\xi _{\mathbf{k}}^{\tau }$'s,%
\begin{eqnarray}
\xi _{\mathbf{k}}^{0} &=&2t_{1}\cos k_{c}+2t_{4}\cos
2k_{c}+2t_{2}s_{x^{2}+y^{2}}(\mathbf{k}),  \notag \\
\xi _{\mathbf{k}}^{8} &=&2t_{1}^{\prime }\cos k_{c}+2t_{4}^{\prime }\cos
2k_{c}+2t_{2}^{\prime }s_{x^{2}+y^{2}}(\mathbf{k}),  \notag \\
\xi _{\mathbf{k}}^{5} &=&Ap_{x}(\mathbf{k}),  \notag \\
\xi _{\mathbf{k}}^{7} &=&Ap_{y}(\mathbf{k}),  \label{H02} \\
\xi _{\mathbf{k}}^{1} &=&Bd_{xy}(\mathbf{k}),  \notag \\
\xi _{\mathbf{k}}^{3} &=&Bd_{x^{2}-y^{2}}(\mathbf{k}),  \notag \\
\xi _{\mathbf{k}}^{4} &=&Cd_{x^{2}-y^{2}}(\mathbf{k}),  \notag \\
\xi _{\mathbf{k}}^{6} &=&-Cd_{xy}(\mathbf{k}),  \notag
\end{eqnarray}%
where $A=(t_{3}+t_{5}\cos k_{c}),B=(\tilde{t}_{3}+\tilde{t}_{5}\cos
k_{c}),C=(t_{3}^{\prime }+t_{5}^{\prime }\cos k_{c})$. By fitting the DFT
band structure of K$_{2}$Cr$_{3}$As$_{3}$, we obtain a set of parameters (in
eV) as following: $t_{1}=0.2237$, $t_{4}=-0.0566$, $t_{2}=-0.0036$, $%
t_{1}^{\prime }=-0.0545$, $t_{4}^{\prime }=0.0214$, $t_{2}^{\prime }=0.0065$%
, $t_{3}=0.0240$, $t_{5}=0.0068$, $\tilde{t}_{3}=-0.0114$, $\tilde{t}_{5}=-0.0099$, $t_{3}^{\prime }=-0.0069$, $t_{5}^{\prime }=-0.0059$. We also
obtain the chemical potential $\mu _{chem}=0.2323$ eV. As shown in Fig. \ref{fig.2}, 
the set of parameters well reproduces the Fermi surfaces. They also
reproduce the energy dispersion along high symmetry lines and the density of
states (DOS) near the Fermi level well.

By diagonalizing $H_{0}$ we obtain three energy bands with dispersion $%
\epsilon _{\mathbf{k}\mu }$ and eigen-wavefunctions $\phi _{\mathbf{k}\mu
}^{m}$, hereafter $\mu $ is the band index. The bare susceptibility tensor $%
\hat{\chi}_{0}(\mathbf{q})$ is defined as%
\begin{eqnarray}
\chi _{0}^{mn,m^{\prime }n^{\prime }}(\mathbf{q}) &=&-\frac{1}{N}\sum_{%
\mathbf{k}}\sum_{\mu \nu }\frac{\phi _{\mathbf{k}\mu }^{m}\phi _{\mathbf{k}%
\mu }^{m^{\prime }\ast }\phi _{\mathbf{k+q}\nu }^{n^{\prime }}\phi _{\mathbf{%
k+q}\nu }^{n\ast }}{\epsilon _{\mathbf{k}\mu }-\epsilon _{\mathbf{k+q}\nu }}
\notag \\
&&\times \lbrack f(\epsilon _{\mathbf{k}\mu })-f(\epsilon _{\mathbf{k+q}\nu
})],  \label{Xiq}
\end{eqnarray}%
where $f(\epsilon _{\mathbf{k}\mu })$ is the Fermion occupation number, $N$
is the number of unit cells. Note that $\hat{\chi}_{0}(\mathbf{q})$ can be
also written as a $9\times 9$ matrix in the two-body orbital space spanned
by the basis $\left\{ \left\vert m\right\rangle \otimes \left\vert
n\right\rangle :m,n=1,2,3\right\} $.

\emph{Interaction:} Now we consider the electron-electron interaction. In
the spirit of Hubbard approximation, i.e. retaining only the intra-unit-cell
terms, we obtain the interacting Hamiltonian which respects $D_{3h}$
symmetry (see Appendix A for details),
\begin{widetext}
\begin{eqnarray}
H_{int} &=&\frac{1}{2}\sum\limits_{i}\left\{
\sum_{m=1}^{2}\sum\limits_{\sigma }[U_{1}c_{im\sigma }^{\dag }c_{im\bar{%
\sigma}}^{\dag }c_{im\bar{\sigma}}c_{im\sigma }+Jc_{im\sigma }^{\dag }c_{im%
\bar{\sigma}}^{\dag }c_{i\bar{m}\bar{\sigma}}c_{i\bar{m}\sigma }+J^{\prime
}(c_{im\sigma }^{\dag }c_{im\bar{\sigma}}^{\dag }c_{i3\bar{\sigma}%
}c_{i3\sigma }+h.c.)]\right. \notag \\
&&+\sum_{m=1}^{2}\sum_{\sigma \sigma ^{\prime }}(U_{2}c_{im\sigma }^{\dag
}c_{i\bar{m}\sigma ^{\prime }}^{\dag }c_{i\bar{m}\sigma ^{\prime
}}c_{im\sigma }+Jc_{im\sigma }^{\dag }c_{i\bar{m}\sigma ^{\prime }}^{\dag
}c_{im\sigma ^{\prime }}c_{i\bar{m}\sigma }+U_{2}^{\prime }c_{im\sigma
}^{\dag }c_{i3\sigma ^{\prime }}^{\dag }c_{i3\sigma ^{\prime }}c_{im\sigma
}+J^{\prime }c_{im\sigma }^{\dag }c_{i3\sigma ^{\prime }}^{\dag }c_{im\sigma
^{\prime }}c_{i3\sigma }) \notag \\
&&\left. +\sum_{\sigma }U_{1}^{\prime }c_{i3\sigma }^{\dag }c_{i3\bar{\sigma}%
}^{\dag }c_{i3\bar{\sigma}}c_{i3\sigma }\right\} , \label{Hint}
\end{eqnarray}
\end{widetext}where $\bar{m}$ is the opposite orbital to $m$ within $%
E^{\prime }$ representation ($\bar{1}=2$ and $\bar{2}=1$), and $\bar{\sigma}$
is the opposite spin to $\sigma $. Here $U_{1}$, $U_{2}$ and $J$ are
intra-orbital repulsion, inter-orbital repulsion and Hund's coupling for the
two $E^{\prime }$ states, which satisfy that $U_{1}=U_{2}+2J$. $%
U_{1}^{\prime }$ is the intra-orbital repulsion for the $A_{1}^{\prime }$
state, and $J^{\prime }$ involves one of $E^{\prime }$ states and the $%
A_{1}^{\prime }$ state.

By Fourier transformation, we have%
\begin{eqnarray}
H_{int} &=&\frac{1}{2}\sum_{\mathbf{kk}^{\prime }\mathbf{q}%
}\sum_{mnm^{\prime }n^{\prime }}\sum_{\sigma \sigma ^{\prime }}\Gamma
_{0}^{mn,m^{\prime }n^{\prime }}(\mathbf{k},\mathbf{k}-\mathbf{q};\mathbf{k}%
^{\prime },\mathbf{k}^{\prime }+\mathbf{q})  \notag \\
&&\times c_{\mathbf{k}m\sigma }^{\dag }c_{\mathbf{k}^{\prime }m^{\prime
}\sigma ^{\prime }}^{\dag }c_{\mathbf{k}^{\prime }\mathbf{+q}n^{\prime
}\sigma ^{\prime }}c_{\mathbf{k-q}n\sigma },  \label{Hintk}
\end{eqnarray}%
where the bare vertex function $\Gamma _{0}^{mn,m^{\prime }n^{\prime }}(%
\mathbf{k},\mathbf{k}-\mathbf{q};\mathbf{k}^{\prime },\mathbf{k}^{\prime }+%
\mathbf{q})$ can be written as a matrix in the two-body orbital space,%
\begin{equation}
\hat{\Gamma}_{0}(\mathbf{k},\mathbf{k}-\mathbf{q};\mathbf{k}^{\prime },%
\mathbf{k}^{\prime }+\mathbf{q})=(1-\delta _{\sigma \sigma ^{\prime }})\hat{%
\Gamma}_{s}+\delta _{\sigma \sigma ^{\prime }}\hat{\Gamma}_{t},
\label{Gamma0}
\end{equation}%
where $\hat{\Gamma}_{t}$ is bare vetex in the equal-spin channel and $\hat{%
\Gamma}_{s}$ is in the opposite-spin channel. The two-body orbital space can
be decomposited to subspaces $\left\{ \left\vert 11\right\rangle ,\left\vert
22\right\rangle ,\left\vert 33\right\rangle \right\} \oplus \{\left\vert
12\right\rangle ,\left\vert 21\right\rangle \}\oplus \{\left\vert
13\right\rangle ,\left\vert 13\right\rangle \}\oplus \{\left\vert
23\right\rangle ,\left\vert 32\right\rangle \}$. Thus, $\hat{\Gamma}_{s}$
and $\hat{\Gamma}_{t}$ are block diagonal in this set of basis, 
\begin{subequations}
\label{GammaST}
\begin{eqnarray}
\hat{\Gamma}_{s} &=&\left( 
\begin{array}{ccc}
U_{1} & U_{2} & U_{2}^{\prime } \\ 
U_{2} & U_{1} & U_{2}^{\prime } \\ 
U_{2}^{\prime } & U_{2}^{\prime } & U_{1}^{\prime }%
\end{array}%
\right) \oplus J\Pi \oplus J^{\prime }\Pi \oplus J^{\prime }\Pi ,
\label{GammaS} \\
\hat{\Gamma}_{t} &=&\left( 
\begin{array}{ccc}
0 & U_{2} & U_{2}^{\prime } \\ 
U_{2} & 0 & U_{2}^{\prime } \\ 
U_{2}^{\prime } & U_{2}^{\prime } & 0%
\end{array}%
\right) \oplus J\sigma _{1}\oplus J^{\prime }\sigma _{1}\oplus J^{\prime
}\sigma _{1},  \label{GammaT}
\end{eqnarray}
where $\sigma _{1}$ is Pauli matrix and $\Pi =1+\sigma _{1}$ is a $2\times 2$
matrix.

\emph{Effective pairing interaction.} For weak coupling, the full vertex
function $\Gamma _{\sigma \sigma ^{\prime }}^{mn,m^{\prime }n^{\prime }}(%
\mathbf{k},\mathbf{k}-\mathbf{q};\mathbf{k}^{\prime },\mathbf{k}^{\prime }+%
\mathbf{q})$ can be evaluated diagrammatically, for instance, through the
random phase approximation (RPA). To study the SC pairing, we only need to
keep the vertices in pairing channels with $\mathbf{k}^{\prime }=-\mathbf{k}$%
, say, $\hat{\Gamma}_{\sigma \sigma ^{\prime }}(\mathbf{k},\mathbf{k}%
^{\prime };-\mathbf{k},-\mathbf{k}^{\prime })\equiv \hat{V}_{\sigma \sigma }(%
\mathbf{k},\mathbf{k}^{\prime })$. Therefore, $\hat{V}_{\sigma \sigma }(%
\mathbf{k},\mathbf{k}^{\prime })=(1-\delta _{\sigma \sigma ^{\prime }})\hat{V%
}_{s}(\mathbf{k},\mathbf{k}^{\prime })+\delta _{\sigma \sigma ^{\prime }}%
\hat{V}_{t}(\mathbf{k},\mathbf{k}^{\prime })$ serves as an effective pairing
interaction to study superconductivity instability.

In study of the cuprates, Scalapino et al. used RPA to calculate $\hat{V}%
_{\sigma \sigma }(\mathbf{k},\mathbf{k}^{\prime })$ in a single Hubbard
model.\cite{Scalapino87} The RPA involves two types of Feynman diagrams in
addition to the bare vertex function $\hat{V}_{0}(\mathbf{k},\mathbf{%
k^{\prime }})$ (see Appendix C for details). One contains the bubble diagrams, and the
other contains the ladder diagrams with Cooperon. Here we generalize the
calculation in Ref.\cite{Scalapino87} to the multi-orbital case. The
effective pairing interaction $\hat{V}_{s(t)}(\mathbf{k},\mathbf{k}^{\prime
})$ from the bubble diagrams is 
\end{subequations}
\begin{eqnarray}
\hat{V}_{s(t)}^{bub}(\mathbf{k},\mathbf{k}^{\prime }) &=&\frac{1}{2}\left\{ (%
\hat{\Gamma}_{t}+\hat{\Gamma}_{s})[1+\hat{\chi}_{0}(\mathbf{q})(\hat{\Gamma}%
_{t}+\hat{\Gamma}_{s})]^{-1}\right.  \notag \\
&&\left. \mp (\hat{\Gamma}_{t}-\hat{\Gamma}_{s})[1+\hat{\chi}_{0}(\mathbf{q}%
)(\hat{\Gamma}_{t}-\hat{\Gamma}_{s})]^{-1}\right\} ,  \label{Vb}
\end{eqnarray}%
where $\mp$ takes $-$ for $s$ and $+$ for $t$. The effective pairing from
the ladder diagrams is%
\begin{equation}
\check{V}_{s(t)}^{lad}(\mathbf{k},\mathbf{k}^{\prime })=\check{\Gamma}_{s(t)}%
\hat{\chi}_{0}(\mathbf{p})\check{\Gamma}_{s(t)}[1-\hat{\chi}_{0}(\mathbf{p})%
\check{\Gamma}_{s(t)}]^{-1},  \label{Vl}
\end{equation}%
where $\mathbf{p=k}+\mathbf{k}^{\prime }$ and $\mathbf{q=k}-\mathbf{k}%
^{\prime }$ (See Appendix C for details). For the notation, matrix $\check{A}$ is
related to matrix $\hat{A}$ via the following relation,%
\begin{equation*}
\check{A}_{mn,m^{\prime }n^{\prime }}=\hat{A}_{nn^{\prime },m^{\prime }m}.
\end{equation*}%
We can also project effective pairing potential $\hat{V}_{\sigma \sigma
^{\prime }}(\mathbf{k},\mathbf{k}^{\prime })$ into three single particle
bands through%
\begin{equation}
V_{\sigma \sigma ^{\prime }}^{\mu \nu }(\mathbf{k},\mathbf{k}^{\prime
})=\sum_{mnm^{\prime }n^{\prime }}\phi _{\mathbf{k}\mu }^{m\ast }\phi _{-%
\mathbf{k}\mu }^{m^{\prime }\ast }V_{\sigma \sigma ^{\prime }}^{mn,m^{\prime
}n^{\prime }}(\mathbf{k},\mathbf{k}^{\prime })\phi _{\mathbf{k}^{\prime }\nu
}^{n}\phi _{-\mathbf{k}^{\prime }\nu }^{n^{\prime }},  \label{Veff}
\end{equation}%
where $\phi _{\mathbf{k}\mu }^{m}$ is the single particle eigenwavefunction
in the $\mu $-band.

\emph{Superconducting pairing instability.} Because the three Fermi surfaces
have different shapes, $\mathbf{k}$ and $-\mathbf{k}$ are always in the same
band for $\mathbf{k}$ near the Fermi surfaces. For weak coupling, we only
consider intra-band pairing ($\mathbf{k}$ to $-\mathbf{k}$). A single band
gap function can be written as%
\begin{equation}
\Delta \left( \mathbf{k}\right) =i\left[ \sigma _{0}\psi \left( \mathbf{k}%
\right) +\mathbf{\sigma }\cdot \mathbf{d}\left( \mathbf{k}\right) \right]
\sigma _{2},  \label{Deltak}
\end{equation}%
where $\psi \left( \mathbf{k}\right) =\psi \left( -\mathbf{k}\right) $ is
the spin-singlet gap function, and the $d$-vector $\mathbf{d}\left( \mathbf{k%
}\right) =-\mathbf{d}\left( -\mathbf{k}\right) $ describes the spin-triplet
pairing, $\sigma _{0}$ is the unit matrix, $\sigma _{1,2,3}$ are Pauli
matrice. To measure the intra-band SC pairing instability, we follow
Scalapino \textit{et al.,}\cite{Scalapino87} and introduce a dimensionless
coupling constant 
\begin{equation*}
\Lambda _{\mu }=-\frac{\int \frac{dS_{\mathbf{k}\mu }}{|\mathbf{v}_{\mathbf{k%
}\mu }|}\int \frac{dS_{\mathbf{k}^{\prime }\mu }}{|\mathbf{v}_{\mathbf{k}%
^{\prime }\mu }|}\sum_{\sigma \sigma ^{\prime }}g_{\sigma \sigma ^{\prime
}}^{\mu }(\mathbf{k})^{\ast }V_{\sigma \sigma ^{\prime }}^{\mu \mu }(\mathbf{%
k},\mathbf{k}^{\prime })g_{\sigma \sigma ^{\prime }}^{\mu }(\mathbf{k}%
^{\prime })}{(2\pi )^{3}\int \frac{dS_{\mathbf{k}\mu }}{|\mathbf{v}_{\mathbf{%
k}\mu }|}\sum_{\sigma \sigma ^{\prime }}|g_{\sigma \sigma ^{\prime }}^{\mu }(%
\mathbf{k})|^{2}},
\end{equation*}%
where $\int dS_{\mathbf{k}\mu }$ is the integration over the $\mu $-band
Fermi surface and $\mathbf{v}_{\mathbf{k}\mu }$ is the Fermi velocity, the $%
\mu $-band form factor $g_{\sigma \sigma ^{\prime }}^{\mu }(\mathbf{k}%
)\propto \Delta _{\sigma \sigma ^{\prime }}(\mathbf{k})$.

In the absence of the spin-orbit coupling, the spin-singlet component $\psi
\left( \mathbf{k}\right) $ will not mix with the spin-triplet component $%
\mathbf{d}\left( \mathbf{k}\right) $ in Eq.(\ref{Deltak}). The possible
single-band SC gap functions on hexagonal lattice are listed in Table \ref%
{gap} up to the first and second NN bonds\cite{Agterberg}.

\begin{figure}[tbph]
\includegraphics[width=8.0cm]{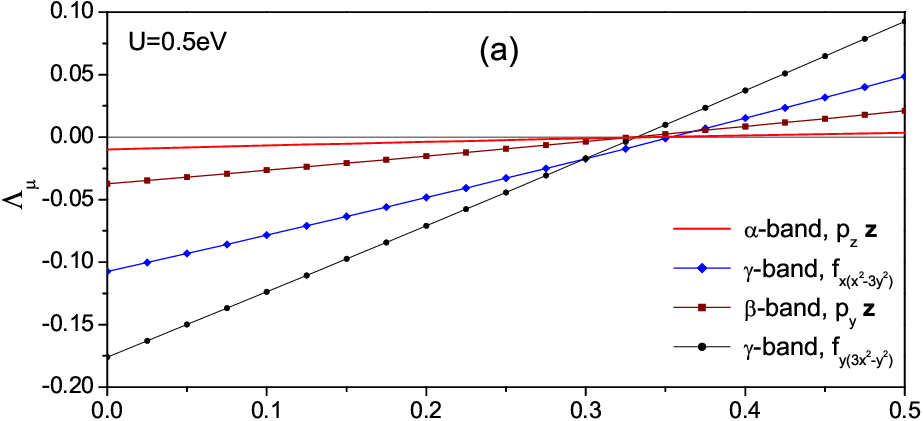} %
\includegraphics[width=8.0cm]{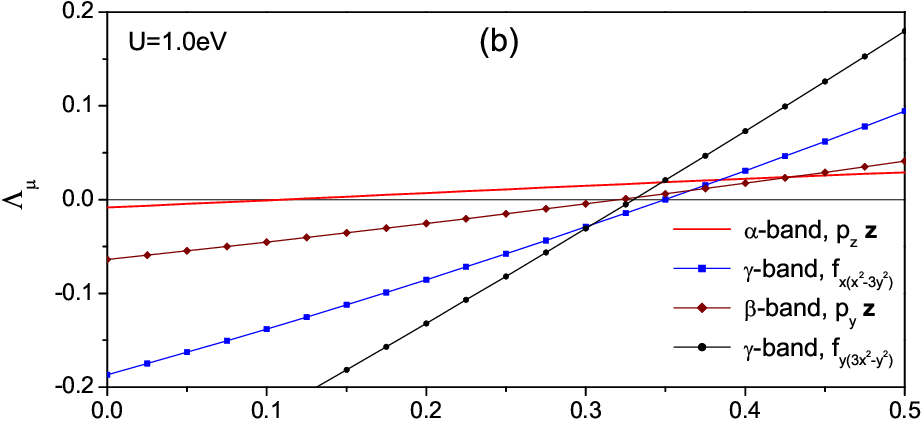} %
\includegraphics[width=8.0cm]{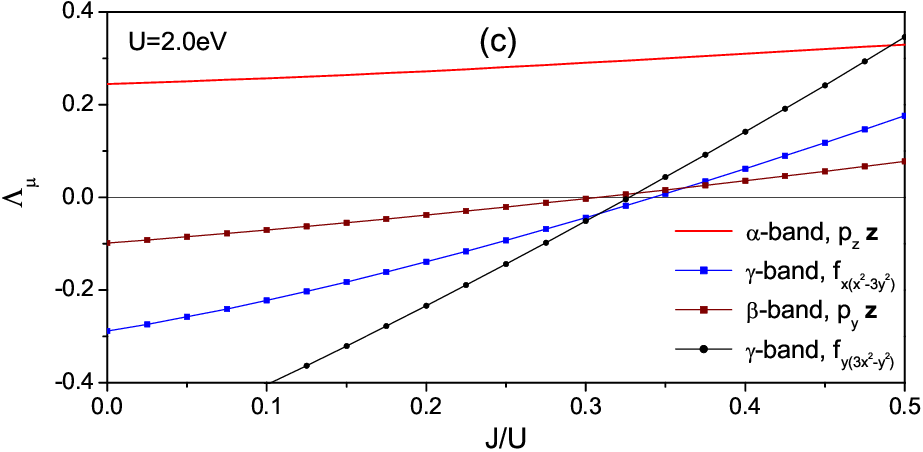}
\caption{ The dimensionless pair coupling constant $\Lambda_{\protect\mu }$
vs $J/U$ for (a) $U=0.5$eV, (b) $U=1.0$eV and (c) $U=2.0$eV. Here we have
set $U_{1}=U_{1}^{\prime }=U$, $U_{2}^{\prime }=U_{2}$ and $J^{\prime }=J$.
All the dominated states are spin-triplet pairing states. (a) $U=0.5$eV, $%
f_{y(3x^{2}-y^{2})}$ appear and dominate when $J/U>1/3$. (b) $U=1.0 $eV, $%
p_{z}\hat{z}$ state and $f_{y(3x^{2}-y^{2})}$ compete against each other.
(c) $U=2.0$eV, $p_{z}\hat{z}$ state dominates over most $J/U$. }
\label{fig.3}
\end{figure}

\begin{table}[tbph]
\caption{Superconducting gap functions $\protect\psi (\mathbf{k}) $ and $%
\mathbf{d}(\mathbf{k})$ on hexagonal lattice. The gap functions are
intra-band and are classified according to $D_{3h}$ group irreducible
representation $\Gamma $. The functions $s_{x^{2}+y^{2}}$, $p_{x}$, $p_{y}$, 
$d_{x^{2}-y^{2}}$, and $d_{xy}$ have been defined in Eq.(\protect\ref{HF1}).
Other functions are defined as follows, $p_{z}=\sin {k_{c}}$, $%
d_{z^{2}}=\cos {k_{c}}$, $f_{x(x^{2}-3y^{2})}=\sin {k_{a}}+\sin {k_{b}}-\sin 
{(k_{a}+k_{b})}$, $f_{y(3x^{2}-y^{2})}=\sin {(2k_{a}+k_{b})}-\sin {%
(k_{a}+2k_{b})}-\sin {(k_{a}-k_{b})}$.}
\label{gap}%
\begin{tabular}{|c|c|c|}
\hline
$\Gamma $ & spin-singlet $\psi (\mathbf{k})$ & spin-triplet $\mathbf{d}(%
\mathbf{k})$ \\ \hline
$A_{1}^{\prime }$ & $1,s_{x^{2}+y^{2}},d_{z^{2}}$ & $f_{y(3x^{2}-y^{2})}\hat{%
z}$ \\ \hline
$A_{2}^{\prime }$ &  & $f_{x(x^{2}-3y^{2})}\hat{z}$ \\ \hline
$E^{\prime }$ & $(d_{x^{2}-y^{2}},d_{xy})$ & $(p_{x},p_{y})\hat{z},p_{z}(%
\hat{x},\hat{y})$ \\ \hline
$A_{1}^{\prime \prime }$ & $p_{z}f_{y(3x^{2}-y^{2})}$ & $p_{x}\hat{x}+p_{y}%
\hat{y},p_{z}\hat{z}$ \\ \hline
$A_{2}^{\prime \prime }$ & $p_{z}f_{x(x^{2}-3y^{2})}$ & $p_{y}\hat{x}-p_{x}%
\hat{y}$ \\ \hline
$E^{\prime \prime }$ & $p_{z}(p_{x},p_{y})$ & $(p_{x}\hat{x}-p_{y}\hat{y}%
,p_{y}\hat{x}+p_{x}\hat{y})$ \\ \hline
\end{tabular}%
\end{table}
We next examine all the pairing states in Table \ref{gap} to investigate
which pairing channel will dominate. For simplicity, we set $%
U_{1}=U_{1}^{\prime }=U$, $U_{2}^{\prime }=U_{2}$ and $J^{\prime }=J$ at
first. The relevant $\Lambda _{\mu }$'s are plotted as functions of $J/U$ in
Fig. \ref{fig.3}, and those $\Lambda_{\mu}$ not plotted have either negative
or negligibly small values.

Fig. 3(a) shows the pair coupling constant in various channels for $U=0.5$
eV. Results for $U<0.5$eV are similar. The spin-triplet $f_{y(3x^{2}-y^{2})}$
state in the $\gamma $-band appears at a finite $J/U$ and become dominant
when $J/U>1/3$. The $f_{y(3x^{2}-y^{2})}$ state has a line nodal gap. Fig.
3(c) plots the results at $U=2$ eV, to represent large $U$. In that case,
the spin-triplet $p_{z}\hat{z}$ state at the $\alpha $-band dominates in all
the realistic region of $J/U<0.45$. The $p_{z}\hat{z}$ state has a full gap
at the quasi-1D $\alpha $-band Fermi surface. In the intermediate region,
e.g., $U=1.0$eV, $p_{z}\hat{z}$ and $f_{y(3x^{2}-y^{2})}$ state compete
against each other, $p_{z}\hat{z}$ dominates at small $J/U$, while $%
f_{y(3x^{2}-y^{2})}$ state become strong at large $J/U$. We have also
calculated the pair coupling constant for $J^{\prime }\neq J$ while keeping $%
U_{1}=U_{1}^{\prime }=U$ and $U_{2}^{\prime }=U_{2}$. The results are
similar in a wide parameter region $0.5<J^{\prime }/J<2.0$. Note that at
large $U$ and $J/U$, the SC pairing is found in the $\beta $-band with
pairing symmetry $p_{x}\hat{x}\pm p_{y}\hat{y}$ and $p_{y}\hat{x}\pm p_{x}%
\hat{y}$, which gives point nodal in the gap function. However, $\Lambda
_{\mu }$ for these point nodally gapped states are tiny.

Note that the divation of our tight-binding model from DFT results in some
details should not change the above statements qualitatively. Because it is
the Fermi surface shape and DOS near the Fermi level that determine $\Lambda
_{\mu }$ and pairing symmetry in weak coupling.

\emph{Discussions and Summary} We would like to mention some issues which
have not been addressed in this paper but may be interesting for future
work. (i) We neglect spin-orbit coupling in the theory for simplicity.
Actually the $D_{3h}$ lattice breaks inversion symmetry, and the spin-orbit
coupling may mix spin singlet and triplet states within the same $D_{3h}$
irreducible representation in Table \ref{gap}. (ii) Since molecular orbitals
are more extended than atomic orbitals, $U$ and $J$ are estimated to be
smaller than or comparable to the bandwidth. In this paper we start with
weak coupling to study SC instability. However, A$_{2}$Cr$_{3}$As$_{3}$
lattice is quasi-1D, an alternative approach would be to model the system as
a coupled Tomonaga-Luttinger liquid. (iii) Although we only consider the
intra-band pairing states in this paper, the inter-band pair scatterings of
Cooper pairs are included in the RPA calculation, since the multi-orbital
susceptibility tensor is used and both particle-hole channels (bubble
diagrams) and particle-particle channels (ladder diagrams) are included in
the RPA.

We would like to make comments on the on-site approximation in the Hubbard
model too. Although the three molecular orbitals are more extended in
comparison with the atomic orbitals, these molecular orbitals are quite
localized. The inter-unit-cell electron interaction is a small fraction of
the intra-unit-cell (see Appendix \ref{app:WF}). Therefore the inter-cell
interaction should play a less important role in our analysis.

The spin-triplet pairing states in our study originates from the Hund's
coupling. The Hund's coupling favors spin-triplet states of two different
molecular orbitals on the same molecule, and Cooper pairs can be formed in
the spin-triplet channel. That's the reason why the triplet pairing channels
get instability when the Hund's coupling $J$ is larger than the
inter-orbital repulsion $U_2$ in the weak coupling scenario\cite{LeeWen08}.
Moreover, the DOS at Fermi level at the 3D $\gamma$-band is much
larger than those at two quasi-1D bands (3 to 5 times), it is expected that
the superconducting instability will dominate at the 3D $\gamma$-band at
small $U$.

In summary, we have proposed a minimal model to study superconductivity in A$%
_{2}$Cr$_{3}$As$_{3}$ (A=K,Rb,Cs), which involves three molecular orbital
states in each unit cell. With the help of symmetry, we have deduced a
tight-binding model with 3 molecular orbitals for the system, which compares
well with the results of the density functional theory. We have derived
effective pairing interactions within the RPA, and found that the dominant
pairing channel is always spin-triplet. For small $U$, a spin-triplet state
with line nodal gap, $f_{y(3x^{2}-y^{2})}$, at the 3D $\gamma $-band will
dominate at moderate Hund's coupling. While for large $U$, a spin-triplet
fully gapped state, $p_{z}\hat{z}$, will dominate at the quasi-1D $\alpha $%
-band. The state we find at small $U$ appears to be most relevant to the
compound. The pairing state of $f_{y(3x^{2}-y^{2})}$ has nodal lines on
planes $k_{a}=k_{b}$, $k_{a}=-2k_{b}$ and $k_{b}=-2k_{a}$, say, $\Gamma $-$K$%
-$L$-$A$ planes on hexagonal lattices. The $\Gamma $-$K$-$L$-$A$ planes
cross small sections at $\gamma $-Fermi surface. Our theory at small $U$
predicts line zeroes in gap function and appears to be consistent with
existing experiments showing non-BCS gap function and particularly the low
temperature London penetration depth measurement. Our prediction can be
tested in further experiments including angle resolved photoemission
spectroscopy.

\emph{Acknowledgment}
We would like to thank G.H. Cao, Z.A Xu, H.Q. Yuan,
F.L. Ning, J. Zhao and H. Yao and D. F. Agterberg for helpful discussions.
This work is supported in part by National Key R\&D Program of China (No.2016YFA0300202),
National Basic Research Program of China
(No.2014CB921201/2014CB921203), NSFC
(No.11374256/11274269/11274006), NSF of Zhejiang Province (No. LR12A04003).

\appendix

\section{Derive the interacting part for three-orbital Hubbard model}

The electron field operator $\hat{\psi}_{\sigma }(\vec{r})$ can be expanded
in terms of a complete set of Wannier functions,%
\begin{equation}
\hat{\psi}_{\sigma }(\vec{r})=\sum_{m}\sum_{i}w_{im}(\vec{r})c_{im\sigma },
\end{equation}%
where $c_{im\sigma }$ annihilates an electron with orbital $m$ and spin $\sigma $ at lattice site $i$. A generic interacting Hamiltonian is given by%
\begin{eqnarray}
H_{int}&=&\frac{1}{2}\sum_{\sigma \sigma ^{\prime }}\int \hat{\psi}_{\sigma}^{\dagger }(\vec{r}_{1})\hat{\psi}_{\sigma ^{\prime }}^{\dagger }(\vec{r}_{2})V(\vec{r}_{1}-\vec{r}_{2}) \nonumber\\
&&\times\hat{\psi}_{\sigma ^{\prime }}(\vec{r}_{2})\hat{\psi}_{\sigma }(\vec{r}_{1})d\vec{r}_{1}d\vec{r}_{2},
\end{eqnarray}
where $V_{ext}(\vec{r})$ is external periodic potential, and $V(\vec{r}_{1}-\vec{r}_{2})$ is the screened Coulomb interaction. $H_{0}$ can be written as
a tight-binding model and we shall focus on the interacting part $H_{int}$.
In the spirit of a Hubbard type approximation, i.e. retaining only the terms
on the same lattice site, we have%
\begin{eqnarray}
H_{int}&=&\frac{1}{2}\sum_{i}\sum_{mnm^{\prime }n^{\prime }}\sum_{\sigma\sigma ^{\prime }}V_{mm^{\prime },n^{\prime }n}\nonumber\\
&& \times c_{im\sigma }^{\dag}c_{im^{\prime }\sigma ^{\prime }}^{\dag }c_{in^{\prime }\sigma ^{\prime}}c_{in\sigma },
\end{eqnarray}%
where%
\begin{eqnarray}
V_{mm^{\prime },n^{\prime }n}&=&\int w_{im}^{\ast }(\vec{r}_{1})w_{im^{\prime}}^{\ast }(\vec{r}_{2})V(\vec{r}_{1}-\vec{r}_{2}) \nonumber\\
&&\times w_{in^{\prime }}(\vec{r}_{2})w_{in}(\vec{r}_{1})d\vec{r}_{1}d\vec{r}_{2}.
\end{eqnarray}
Let's consider nonvanishing $V_{mm^{\prime },n^{\prime }n}$ according to $D_{3h}$ group symmetry. Since the interaction $V(\vec{r}_{1}-\vec{r}_{2})=V(|\vec{r}_{1}-\vec{r}_{2}|)$ 
respects the full point group symmetry and belong to the $A_{1}$ representation. Wether a $V_{mm^{\prime },n^{\prime }n}$ vanishes can be determined by the Clebsch-Gordan coefficients of $D_{3h}$ group.

Firstly, for $m,n=1,2$, the nonzero $V_{mm^{\prime },n^{\prime }n}$ can be
written in terms of the integrals explicitly, 
\begin{eqnarray*}
V_{mm,mm} &=&\int |w_{im}(\vec{r}_{1})|^{2}V(\vec{r}_{1}-\vec{r}_{2})|w_{im}(\vec{r}_{2})|^{2}d\vec{r}_{1}d\vec{r}_{2}\\
&=&U_{1}, \\
V_{m\bar{m},\bar{m}m} &=&\int |w_{im}(\vec{r}_{1})|^{2}V(\vec{r}_{1}-\vec{r}_{2})|w_{i\bar{m}}(\vec{r}_{2})|^{2}d\vec{r}_{1}d\vec{r}_{2}\\
&=&U_{2}, \\
V_{m\bar{m},m\bar{m}} &=&\int w_{im}^{\ast }(\vec{r}_{1})w_{i\bar{m}}(\vec{r}_{1})V(\vec{r}_{1}-\vec{r}_{2})\\
&&\times w_{i\bar{m}}^{\ast }(\vec{r}_{2})w_{im}(\vec{r}_{2})d\vec{r}_{1}d\vec{r}_{2}=J, \\
V_{mm,\bar{m}\bar{m}} &=&\int w_{im}^{\ast }(\vec{r}_{1})w_{i\bar{m}}(\vec{r}_{1})V(\vec{r}_{1}-\vec{r}_{2})\\
&&\times w_{i\bar{m}}(\vec{r}_{2})w_{im}^{\ast }(\vec{r}_{2})d\vec{r}_{1}d\vec{r}_{2}=J^{\ast }.
\end{eqnarray*}
where $U_{1}$ is the intra-orbital interaction and $U_{2}$ is the inter-orbital interaction. If one chooses the Wannier function to be real,
then $J^{\ast }=J$, and $J$ is the Hund's exchange energy.

Secondly, we can choose that $\left\vert 1\right\rangle $ transfers as $x$
and $\left\vert 2\right\rangle $ transfers as $y$ under $D_{3h}$ symmetry
operations. Note that $V(\vec{r}_{1}-\vec{r}_{2})$ is invariant not only
under $D_{3h}$ symmetry operations, but also under all the $O(3)$ symmetry
operations. So that $V_{xx,xx}=V_{yy,yy}=U_{1}$. Under a $C_{3}$ rotation
along the $c$-axis, the two Wannier functions $\left\vert 1\right\rangle $
and $\left\vert 2\right\rangle $ (denoted by $x$ and $y$) transfer as%
\begin{eqnarray*}
w_{x} &\rightarrow &\cos \theta w_{x}+\sin \theta w_{y}, \\
w_{y} &\rightarrow &-\sin \theta w_{x}+\cos \theta w_{y},
\end{eqnarray*}%
where $\theta =\frac{2\pi }{3}$. The integral $V_{xx,xx}$ should keep
invariant under this operation. Assuming $w_{x}$ and $w_{y}$ are real, after
straightforward algebra, we have%
\begin{eqnarray*}
U_{1}&=&(\cos ^{4}\theta +\sin ^{4}\theta )U_{1}+2U_{2}\cos ^{2}\theta \sin^{2}\theta \\
&&+4\cos ^{2}\theta \sin ^{2}\theta J,
\end{eqnarray*}%
where the transfermation peroperties under $C_{2}$ are used. From the above,
we have%
\begin{equation*}
U_{1}=U_{2}+2J.
\end{equation*}

Thirdly, we involve the state $\left\vert 3\right\rangle $. For $m,m^{\prime
}=1,2$ (or $x,y$), the relevant nonvanishing terms are given in the
following, 
\begin{eqnarray*}
V_{33,33} &=&\int |w_{i3}(\vec{r}_{1})|^{2}V(\vec{r}_{1}-\vec{r}_{2})|w_{i3}(\vec{r}_{2})|^{2}d\vec{r}_{1}d\vec{r}_{2}\\
&=&U_{1}^{\prime }, \\
V_{3m,m3} &=&\int |w_{im}(\vec{r}_{1})|^{2}V(\vec{r}_{1}-\vec{r}_{2})|w_{i3}(\vec{r}_{2})|^{2}d\vec{r}_{1}d\vec{r}_{2}\\
&=&U_{2}^{\prime }, \\
V_{m3,m3} &=&\int w_{im}^{\ast }(\vec{r}_{1})w_{i3}(\vec{r}_{1})V(\vec{r}_{1}-\vec{r}_{2})\\
&&\times w_{i3}^{\ast }(\vec{r}_{2})w_{im}(\vec{r}_{2})d\vec{r}_{1}d\vec{r}_{2}=J^{\prime }, \\
V_{mm^{\prime },33} &=&\int w_{im}^{\ast }(\vec{r}_{1})w_{i3}(\vec{r}_{1})V(\vec{r}_{1}-\vec{r}_{2})\\
&&\times w_{im^{\prime }}^{\ast }(\vec{r}_{2})w_{i3}(\vec{r}_{2})d\vec{r}_{1}d\vec{r}_{2}=J_{mm^{\prime }}^{\prime \prime }.
\end{eqnarray*}%
When Wannier functions are real, $J_{mm}^{\prime \prime }=J^{\prime }$. For $%
m\neq m^{\prime }$, we set $m=x$ and $m^{\prime }=y$, thus under the $C_{3}$
rotation,%
\begin{eqnarray*}
J_{xx}^{\prime \prime }&=&\cos ^{2}\theta J_{xx}^{\prime \prime }+\sin^{2}\theta J_{yy}^{\prime \prime }+2\cos \theta \sin \theta J_{xy}^{\prime\prime }\\
&=&J_{xx}^{\prime \prime }+2\cos \theta \sin \theta J_{xy}^{\prime\prime }.
\end{eqnarray*}%
The above gives rise to%
\begin{equation*}
J_{xy}^{\prime \prime }=0.
\end{equation*}%

Thus, we obtain $H_{int}$ in Eq.~(\ref{Hint}) in the main text.

\section{Wannier functions from density functional theory}\label{app:WF}

\begin{figure}[tbph]
\begin{center}
\includegraphics[width=8.0cm]{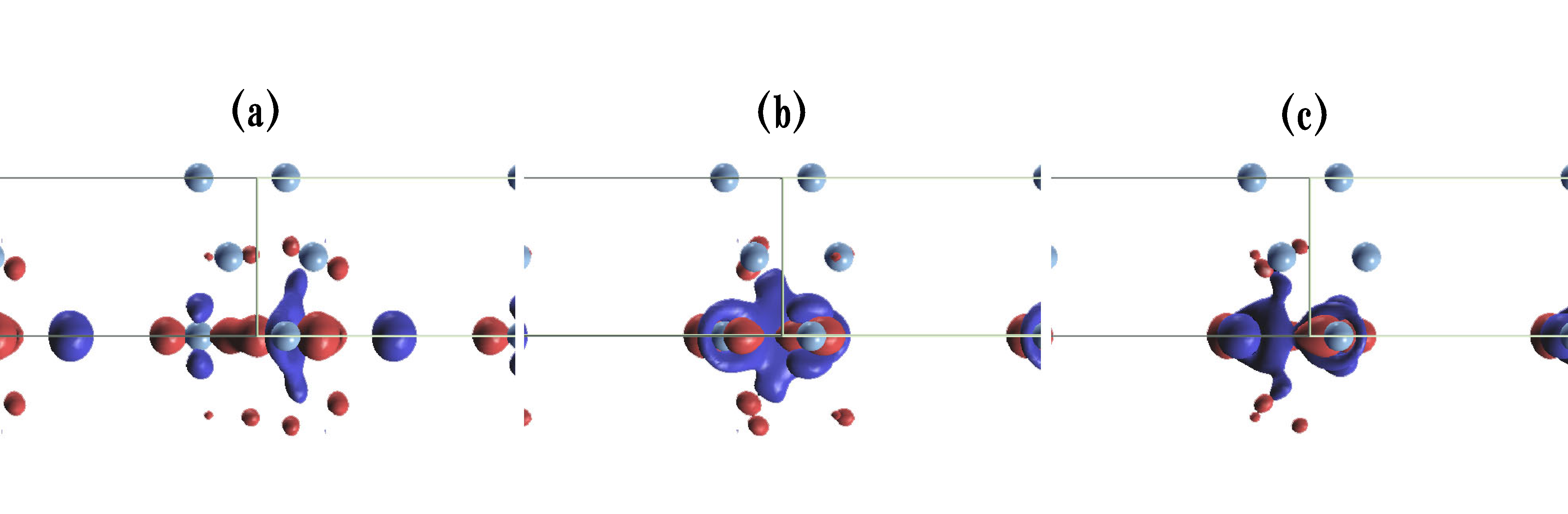} %
\includegraphics[width=8.0cm]{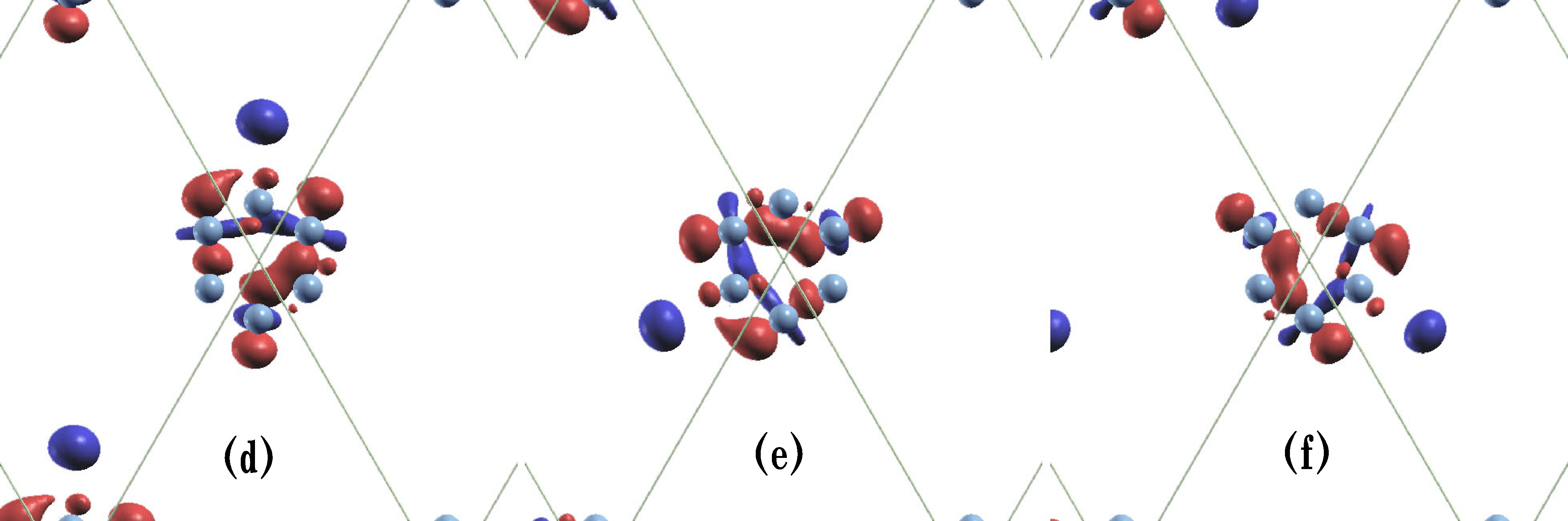}
\end{center}
\caption{ Three Wannier functions from DFT calculation. (a)(b)(c) Side view
of three Wannier functions. (d)(e)(f) Top view of three Wannier functions. }
\label{fig.Wan}
\end{figure}

To estimate the order of magnitude of intra-unit-cell $U$ and
inter-unit-cell interaction $V$, we extract the Wannier functions for the
three energy bands crossing Fermi levels from the DFT calculation. It should
be noted these three Wannier functions $\tilde{w}_{im=1,2,3}$ are not the
molecular orbitals $w_{im=1,2,3}$ used in our tight-binding model. Indeed $%
\tilde{w}_{im}$ are some linear combinations of $w_{im}$. So that we can not
calculate $U_{1},U_{2},U_{1}^{\prime },U_{2}^{\prime }$ and $J,J^{\prime }$
from the Wannier functions $\tilde{w}_{im}$ directly. However, we can
roughly estimate the order of magnitude of $U$ and $V$ from $\tilde{w}_{im}$%
. To do this, we use unscreened Coloumb interaction $e^{2}/r$ to estimate
the on-site and nearest-neighboring interaction $U$ and $V$,%
\begin{eqnarray*}
U_{mm^{\prime }} &=&\int |\tilde{w}_{im}(\vec{r}_{1})|^{2}\frac{e^{2}}{|\vec{%
r}_{1}-\vec{r}_{1}|}|\tilde{w}_{im^{\prime }}(\vec{r}_{2})|^{2}d\vec{r}_{1}d%
\vec{r}_{2}, \\
V_{mm^{\prime }} &=&\int |\tilde{w}_{im}(\vec{r}_{1})|^{2}\frac{e^{2}}{|\vec{%
r}_{1}-\vec{r}_{1}|}|\tilde{w}_{i+\hat{z},m^{\prime }}(\vec{r}_{2})|^{2}d%
\vec{r}_{1}d\vec{r}_{2},
\end{eqnarray*}%
where $i+\hat{z}$ denote the neigboring unit cell to the unit cell $i$ along
the $c$-axis.

It should be noted that the real interaction should be smaller due to
screening effect. As mentioned, $U_{mm^{\prime }}$ and $V_{mm^{\prime }}$
will be reduced due to secreening effect. The present values are used to
estimate order of magnitde and the ratio $V/U$ only.

\begin{table}[tbph]
\caption{$U_{mm^{\prime }}$ and $V_{mm^{\prime }}$, which should be reduced
due to secreening.}
\label{UV}
\begin{center}
\begin{tabular}{|c|c|c|c|}
\hline
$m$ & $m^{\prime} $ & $U_{mm^{\prime }}($eV$)$ & $V_{mm^{\prime }}($eV$)$ \\ 
\hline
$1$ & $1$ & 2.195 & 0.169 \\ \hline
$2$ & $2$ & 2.128 & 0.161 \\ \hline
$3$ & $3$ & 2.143 & 0.155 \\ \hline
$1$ & $2$ & 2.082 & 0.163 \\ \hline
$2$ & $3$ & 2.068 & 0.154 \\ \hline
$1$ & $3$ & 2.107 & 0.159 \\ \hline
\end{tabular}
\end{center}
\end{table}

We can also estimate the spreading of Wannier functions along the $c$-axis
and $ab$-plane respectively, which are defined by%
\begin{eqnarray*}
r_{\parallel m}^{2} &=&\int z^{2}|\tilde{w}_{im}(\vec{r})|^{2}d\vec{r}, \\
r_{\perp m}^{2} &=&\int (x^{2}+y^{2})|\tilde{w}_{im}(\vec{r})|^{2}d\vec{r}.
\end{eqnarray*}%
It is estimated that $r_{\parallel }=1.7$ \r{A} and $r_{\perp }=4.8$ \r{A}.
For the Cr atom, the 3$d$ wavefunction spreading is about $r_{0}=1.6$ \r{A}.
The atomic on-site interaction $U$ for Cr is about $2$ eV. Then we can
estimate the interaction $U$ for a Cr$_{6}$As$_{6}$ cluster as less than $1$
eV.

\section{RPA for effective pairing interaction}

\begin{figure}[tbph]
\includegraphics[width=8.4cm]{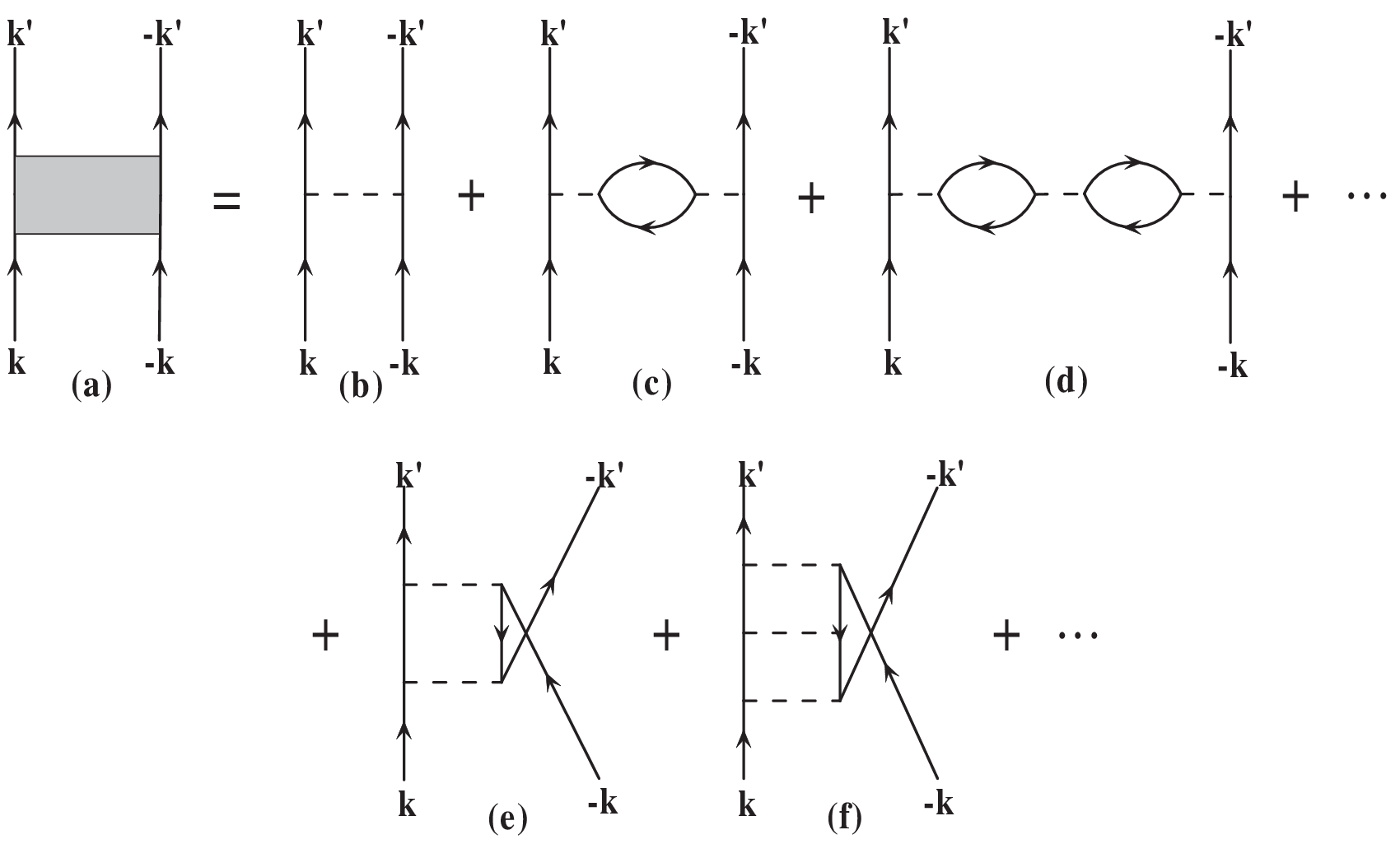}
\caption{ Feynman diagram for RPA calculation of effective pairing
interaction $V(\mathbf{k},\mathbf{k^{\prime}})$. (a) Effective vertex
function $V(\mathbf{k},\mathbf{k^{\prime}})$. (b) Bare vertex function $V_0(%
\mathbf{k},\mathbf{k^{\prime}})$. (c) and (d) Bubble diagrams. (e) and (f)
Ladder digarams with Cooperon. }
\label{fig.RPA}
\end{figure}

We shall generalized the calculation by Scalapino et al. to the
multi-orbital case. Diagramatically, RPA contains two types Feynman diagrams
beside the bare vertex function $V_{0}(\mathbf{k},\mathbf{k^{\prime }})$ as
shown in Fig. \ref{fig.RPA}. One contains the bubble diagrams, the other
contains the ladder diagrams with Cooperon.

For the single orbital case, the RPA effective pairing interaction for
spin-singlet and spin-triplet channels are%
\begin{eqnarray*}
V_{s}(\mathbf{k},\mathbf{k}^{\prime }) &=&\frac{U}{1-U^{2}\chi _{0}^{2}(%
\mathbf{k}-\mathbf{k}^{\prime })}+\frac{U^{2}\chi _{0}(\mathbf{k}+\mathbf{k}%
^{\prime })}{1-U\chi _{0}(\mathbf{k}+\mathbf{k}^{\prime })}, \\
V_{t}(\mathbf{k},\mathbf{k}^{\prime }) &=&-\frac{U^{2}\chi _{0}(\mathbf{k}-%
\mathbf{k}^{\prime })}{1-U^{2}\chi _{0}^{2}(\mathbf{k}-\mathbf{k}^{\prime })}%
.
\end{eqnarray*}%
Note that Fig. \ref{fig.RPA}(c) is absent in $U_{s}$ and Fig. \ref{fig.RPA}%
(b), (d), (e) and (f) are absent in $V_{t}$ because $\sigma ^{\prime }=\bar{%
\sigma}$ in the $U$ term.

For multi-obtial case, we shall replace the numbers $U$ and $\chi _{0}$ by
the bare vertex functions $\hat{\Gamma}_{s(t)}$ in Eq. (9) and
susceptibility tensor $\hat{\chi}_{0}$ in Eq. (5) in the main text. The
effective pairing interaction $\hat{V}_{s(t)}(\mathbf{k},\mathbf{k}^{\prime
})$ from the bubble diagrams are 
\begin{eqnarray*}
\hat{V}_{s}^{bub}(\mathbf{k},\mathbf{k}^{\prime }) &=&\hat{\Gamma}_{s}-\hat{\Gamma}_{s}\hat{\chi}_{0}(\mathbf{q})\hat{\Gamma}_{t}-\hat{\Gamma}_{t}\hat{\chi}_{0}(\mathbf{q})\hat{\Gamma}_{s} \\
&&+\hat{\Gamma}_{s}\hat{\chi}_{0}(\mathbf{q})\hat{\Gamma}_{t}\hat{\chi}_{0}(\mathbf{q})\hat{\Gamma}_{t}\\
&&+\hat{\Gamma}_{t}\hat{\chi}_{0}(\mathbf{q})\hat{\Gamma}_{s}\hat{\chi}_{0}(\mathbf{q})\hat{\Gamma}_{t}\\
&&+\hat{\Gamma}_{t}\hat{\chi}_{0}(\mathbf{q})\hat{\Gamma}_{t}\hat{\chi}_{0}(\mathbf{q})\hat{\Gamma}_{s}\\
&&+\hat{\Gamma}_{s}\hat{\chi}_{0}(\mathbf{q})\hat{\Gamma}_{s}\hat{\chi}_{0}(\mathbf{q})\hat{\Gamma}_{s}+\cdots \\
&&+\hat{\Gamma}_{s}\hat{\chi}_{0}(\mathbf{p})\hat{\Gamma}_{s}+\hat{\Gamma}_{s}\hat{\chi}_{0}(\mathbf{p})\hat{\Gamma}_{s}\hat{\chi}_{0}(\mathbf{p})\hat{\Gamma}_{s}+\cdots \\
&=&\frac{1}{2}(\hat{\Gamma}_{t}+\hat{\Gamma}_{s})[1+\hat{\chi}_{0}(\mathbf{q})(\hat{\Gamma}_{t}+\hat{\Gamma}_{s})]^{-1}\\
&&-\frac{1}{2}(\hat{\Gamma}_{t}-\hat{\Gamma}_{s})[1+\hat{\chi}_{0}(\mathbf{q})(\hat{\Gamma}_{t}-\hat{\Gamma}_{s})]^{-1}
\end{eqnarray*}
and%
\begin{eqnarray*}
\hat{V}_{t}^{bub}(\mathbf{k},\mathbf{k}^{\prime }) &=&\hat{\Gamma}_{t}-\hat{\Gamma}_{s}\hat{\chi}_{0}(\mathbf{q})\hat{\Gamma}_{s}-\hat{\Gamma}_{t}\hat{\chi}_{0}(\mathbf{q})\hat{\Gamma}_{t} \\
&&+\hat{\Gamma}_{t}\hat{\chi}_{0}(\mathbf{q})\hat{\Gamma}_{t}\hat{\chi}_{0}(\mathbf{q})\hat{\Gamma}_{t}\\
&&+\hat{\Gamma}_{t}\hat{\chi}_{0}(\mathbf{q})\hat{\Gamma}_{s}\hat{\chi}_{0}(\mathbf{q})\hat{\Gamma}_{s}\\
&&+\hat{\Gamma}_{s}\hat{\chi}_{0}(\mathbf{q})\hat{\Gamma}_{t}\hat{\chi}_{0}(\mathbf{q})\hat{\Gamma}_{s}\\
&&+\hat{\Gamma}_{s}\hat{\chi}_{0}(\mathbf{q})\hat{\Gamma}_{s}\hat{\chi}_{0}(\mathbf{q})\hat{\Gamma}_{t}+\cdots \\
&&+\hat{\Gamma}_{t}\hat{\chi}_{0}(\mathbf{p})\hat{\Gamma}_{t}+\hat{\Gamma}_{t}\hat{\chi}_{0}(\mathbf{p})\hat{\Gamma}_{t}\hat{\chi}_{0}(\mathbf{p})\hat{\Gamma}_{t}+\cdots \\
&=&\frac{1}{2}(\hat{\Gamma}_{t}+\hat{\Gamma}_{s})[1+\hat{\chi}_{0}(\mathbf{q})(\hat{\Gamma}_{t}+\hat{\Gamma}_{s})]^{-1}\\
&&+\frac{1}{2}(\hat{\Gamma}_{t}-\hat{\Gamma}_{s})[1+\hat{\chi}_{0}(\mathbf{q})(\hat{\Gamma}_{t}-\hat{\Gamma}_{s})]^{-1},
\end{eqnarray*}%
while those from ladder diagrams are%
\begin{eqnarray*}
\check{V}_{s}^{lad}(\mathbf{k},\mathbf{k}^{\prime }) &=&\check{\Gamma}_{s}%
\hat{\chi}_{0}(\mathbf{p})\check{\Gamma}_{s}+\check{\Gamma}_{s}\hat{\chi}%
_{0}(\mathbf{p})\check{\Gamma}_{s}\hat{\chi}_{0}(\mathbf{p})\check{\Gamma}%
_{s}+\cdots \\
&=&\check{\Gamma}_{s}\hat{\chi}_{0}(\mathbf{p})\check{\Gamma}_{s}[1-\hat{\chi%
}_{0}(\mathbf{p})\check{\Gamma}_{s}]^{-1}
\end{eqnarray*}%
and%
\begin{eqnarray*}
\check{V}_{t}^{lad}(\mathbf{k},\mathbf{k}^{\prime }) &=&\check{\Gamma}_{t}%
\hat{\chi}_{0}(\mathbf{p})\check{\Gamma}_{t}+\check{\Gamma}_{t}\hat{\chi}%
_{0}(\mathbf{p})\check{\Gamma}_{t}\hat{\chi}_{0}(\mathbf{p})\check{\Gamma}%
_{t}+\cdots \\
&=&\check{\Gamma}_{t}\hat{\chi}_{0}(\mathbf{p})\check{\Gamma}_{t}[1-\hat{\chi%
}_{0}(\mathbf{p})\check{\Gamma}_{t}]^{-1},
\end{eqnarray*}%
where $\mathbf{p=k}+\mathbf{k}^{\prime }$ and $\mathbf{q=k}-\mathbf{k}%
^{\prime }$, and the matrix $\check{A}$ is related to matrix $\hat{A}$
through the following relation,%
\begin{equation*}
\check{A}_{mn,m^{\prime }n^{\prime }}=\hat{A}_{nn^{\prime },m^{\prime }m}.
\end{equation*}

\subsection{Matrix summation in the bubble diagrams}

Below we write down the details on summing over the two series of matrice in
the bubble diagrams,%
\begin{eqnarray*}
\hat{A} &=&\hat{\Gamma}_{s}-\hat{\Gamma}_{s}\hat{\chi}_{0}(\mathbf{q})\hat{\Gamma}_{t}-\hat{\Gamma}_{t}\hat{\chi}_{0}(\mathbf{q})\hat{\Gamma}_{s} \\
&&+\hat{\Gamma}_{s}\hat{\chi}_{0}(\mathbf{q})\hat{\Gamma}_{t}\hat{\chi}_{0}(\mathbf{q})\hat{\Gamma}_{t}+\hat{\Gamma}_{t}\hat{\chi}_{0}(\mathbf{q})\hat{\Gamma}_{s}\hat{\chi}_{0}(\mathbf{q})\hat{\Gamma}_{t}\\
&&+\hat{\Gamma}_{t}\hat{\chi}_{0}(\mathbf{q})\hat{\Gamma}_{t}\hat{\chi}_{0}(\mathbf{q})\hat{\Gamma}_{s}+\hat{\Gamma}_{s}\hat{\chi}_{0}(\mathbf{q})\hat{\Gamma}_{s}\hat{\chi}_{0}(\mathbf{q})\hat{\Gamma}_{s}+\cdots
\end{eqnarray*}%
and%
\begin{eqnarray*}
\hat{B} &=&\hat{\Gamma}_{t}-\hat{\Gamma}_{s}\hat{\chi}_{0}(\mathbf{q})\hat{\Gamma}_{s}-\hat{\Gamma}_{t}\hat{\chi}_{0}(\mathbf{q})\hat{\Gamma}_{t} \\
&&+\hat{\Gamma}_{t}\hat{\chi}_{0}(\mathbf{q})\hat{\Gamma}_{t}\hat{\chi}_{0}(\mathbf{q})\hat{\Gamma}_{t}+\hat{\Gamma}_{t}\hat{\chi}_{0}(\mathbf{q})\hat{\Gamma}_{s}\hat{\chi}_{0}(\mathbf{q})\hat{\Gamma}_{s}\\
&&+\hat{\Gamma}_{s}\hat{\chi}_{0}(\mathbf{q})\hat{\Gamma}_{t}\hat{\chi}_{0}(\mathbf{q})\hat{\Gamma}_{s}+\hat{\Gamma}_{s}\hat{\chi}_{0}(\mathbf{q})\hat{\Gamma}_{s}\hat{\chi}_{0}(\mathbf{q})\hat{\Gamma}_{t}+\cdots
\end{eqnarray*}
where every term in $\hat{A}$ cotains odd number of $\hat{\Gamma}_{s}$ and
every term in $\hat{B}$ contains even number of $\hat{\Gamma}_{s}$, and each 
$\hat{\chi}_{0}$ contribute one minus sign. To do this summation, we
multiple $\hat{\chi}_{0}(\mathbf{q})$ at both sides at first, then add them
together or substract one to another. Hence%
\begin{eqnarray*}
\hat{\chi}_{0}(\mathbf{q})\hat{B}+\hat{\chi}_{0}(\mathbf{q})\hat{A} &=&\hat{\chi}_{0}(\mathbf{q})(\hat{\Gamma}_{t}+\hat{\Gamma}_{s})\\
&&\times [1+\hat{\chi}_{0}(\mathbf{q})(\hat{\Gamma}_{t}+\hat{\Gamma}_{s})]^{-1}, \\
\hat{\chi}_{0}(\mathbf{q})\hat{B}-\hat{\chi}_{0}(\mathbf{q})\hat{A} &=&\hat{\chi}_{0}(\mathbf{q})(\hat{\Gamma}_{t}-\hat{\Gamma}_{s})\\
&&\times [1+\hat{\chi}_{0}(\mathbf{q})(\hat{\Gamma}_{t}-\hat{\Gamma}_{s})]^{-1}.
\end{eqnarray*}
Then we obtain that
\begin{eqnarray*}
\hat{A} &=&\frac{1}{2}(\hat{\Gamma}_{t}+\hat{\Gamma}_{s})[1+\hat{\chi}_{0}(\mathbf{q})(\hat{\Gamma}_{t}+\hat{\Gamma}_{s})]^{-1}-\frac{1}{2}(\hat{\Gamma}_{t}\\
&&-\hat{\Gamma}_{s})[1+\hat{\chi}_{0}(\mathbf{q})(\hat{\Gamma}_{t}-\hat{\Gamma}_{s})]^{-1}, \\
\hat{B} &=&\frac{1}{2}(\hat{\Gamma}_{t}+\hat{\Gamma}_{s})[1+\hat{\chi}_{0}(\mathbf{q})(\hat{\Gamma}_{t}+\hat{\Gamma}_{s})]^{-1}+\frac{1}{2}(\hat{\Gamma}_{t}\\
&&-\hat{\Gamma}_{s})[1+\hat{\chi}_{0}(\mathbf{q})(\hat{\Gamma}_{t}-\hat{\Gamma}_{s})]^{-1}.
\end{eqnarray*}


\begin{thebibliography}{99}

\bibitem{Luo14} W. Wu, J. Cheng, K. Matsubayashi, P. Kong, F. Lin, C. Jin,
N. Wang, Y. Uwatoko, and J. Luo, Nat. Commun. 5, 5508 (2014).

\bibitem{BaoK} Jin-Ke Bao, Ji-Yong Liu, Cong-Wei Ma, Zhi-Hao Meng, Zhang-Tu
Tang, Yun-Lei Sun, Hui-Fei Zhai, Hao Jiang, Hua Bai, Chun-Mu Feng, Zhu-An
Xu, Guang-Han Cao, Phys. Rev. X 5, 011013 (2015).

\bibitem{TangRb} Zhang-Tu Tang, Jin-Ke Bao, Yi Liu, Yun-Lei Sun, Abduweli
Ablimit, Hui-Fei Zhai, Hao Jiang, Chun-Mu Feng, Zhu-An Xu, Guang-Han Cao,
Phys. Rev. B 91, 020506(R) (2015).

\bibitem{TangCs} Zhang-Tu Tang, Jin-Ke Bao, Zhen Wang, Hua Bai, Hao Jiang,
Yi Liu, Hui-Fei Zhai, Chun-Mu Feng, Zhu-An Xu, Guang-Han Cao, Science China
Materials, 58(1), 16-10 (2015).

\bibitem{Imai15} H. Z. Zhi, T. Imai, F. L. Ning, Jin-Ke Bao, Guang-Han Cao,
Phys. Rev. Lett. 114, 147004 (2015).

\bibitem{Yuan15} G. M. Pang, M. Smidman, W. B. Jiang, J. K. Bao, Z. F. Weng,
Y. F. Wang, L. Jiao, J. L. Zhang, G. H. Cao, H. Q. Yuan, Phys. Rev. B 91, 220502(R) (2015). 

\bibitem{Cao14} Hao Jiang, Guanghan Cao, Chao Cao, Scientific Reports 5, 16054 (2015).

\bibitem{HuJP15} Xianxin Wu, Congcong Le, Jing Yuan, Heng Fan, Jiangping Hu, Chin. Phys. Lett. 32,057401(2015) 

\bibitem{Canfield15} Tai Kong, Sergey L. Bud'ko, Paul C. Canfield, Phys. Rev. B 91, 020507 (2015) .

\bibitem{Butler} P. H. Butler, \emph{Point Group Symmetry Applications:
Methods and Tables}, Plenum Press, New York (1981).

\bibitem{Scalapino87} D. J. Scalapino, E. Loh, Jr., and J. E. Hirsch, Phys.
Rev. B 35, 6694 (1987).

\bibitem{Agterberg} In private communication with Daniel F. Agterberg, we learnt that he had a smiliar table as ours.

\bibitem{LeeWen08} Patrick A. Lee and Xiao-Gang Wen, Phys. Rev. B 78, 144517 (2008).
\end{thebibliography}
\end{document}